\documentclass{aa501}
\usepackage{graphicx}
\begin{document}  

	\title{Photometric validation of a model independent procedure    
		to extract galaxy clusters\thanks{Based on observation   
collected at the European Southern Observatory, Chile, ESO N$^\circ$   
62.O-0230 and 64.O-0317(A).}}

	\subtitle{}       
  
	\author{E. Puddu\inst{1}\and S. Andreon\inst{1}\and G. Longo  
   		\inst{1,2}\and V. Strazzullo\inst{1}\and M. Paolillo\inst{1,3}
                 \and R.R. Gal\inst{4}  
           	}  
  
  
	\institute{Osservatorio Astronomico di Capodimonte,via Moiariello  
 		   16, I-80131 Napoli, Italy\\  
            	  \and
		   Dipartimento di Scienze Fisiche, Universit\`a Federico II,
		   Napoli, Italy\\  
		  \and
		   Dipartimento di Scienze Fisiche ed Astronomiche, 
		   Universit\`a di Palermo, Napoli, Italy\\
		  \and
                   Johns Hopkins University, Dept. of Physics and Astronomy,  
		   3701 San Martin Dr., Baltimore, MD 21218   
                                     }       
	\date{Received ??; accepted ??}       
	\abstract{  
By means of CCD photometry in three bands (Gunn {\it g, r, i})  
 we investigate the existence of 12 candidate clusters extracted  
via a model independent peak finding algorithm (\cite{memsait})   
from DPOSS data. The derived color-magnitude diagrams  
allow us to confirm the physical nature of 9 of the cluster candidates,  
and to estimate their photometric redshifts.  Of  the other  
candidates, one is a fortuitous detection of a true cluster at $z \sim 0.4$,   
one is a false detection and the last is undecidable on the  basis   
of the available data.   
The accuracy of the photometric redshifts is tested on an additional  
sample of 8 clusters with known spectroscopic redshifts. Photometric
redshifts turn out to be accurate within z$\sim 0.01$ (interquartile range).
	\keywords{methods:data analysis - galaxies:clustering - galaxies:clusters:  
photometric redshift}  
		 }  
  
	\maketitle       
%

\section{INTRODUCTION}   
  
 Clusters of galaxies are the largest virialized structures  in the  
 Universe, and accurate knowledge of their global properties is  
 needed to constrain models of galaxy formation and evolution.\\  
 The first step in this direction requires the construction of a  
 statistically well-defined sample of clusters in the nearby universe   
 to be used as a "local template". Unfortunately, due to historical and   
 observational reasons, and in spite of much effort, existing samples cannot  
be considered ideal. Existing cluster catalogs, in fact, fall into  
 three main categories: i) large catalogs derived from photographic  
 surveys (POSS-I, UKST) by visual inspection and covering wide  
 portions of the sky (\cite{Abell1}; \cite{Abell2}; \cite{Zwi})    
 but missing the needed depth, homogeneity and completeness  
(\cite{Post86}; \cite{Suth88}); ii) catalogs machine extracted   
with objective criteria from photographic plates (cf.  
\cite{DMac86}; \cite{Dal92}; \cite{Lum92}),  
reaching, in some cases, limiting magnitudes fainter than (i), but not covering  
equally wide areas of the sky and so far available only for the Southern  
hemisphere (only UKST plates);  
iii) accurate and deeper catalogs usually derived from CCD data and   
selected on the basis of objective criteria but covering   
much smaller regions of the sky and containing only a small number of objects   
(cf. \cite{Post96}; \cite{Olsen99}).
Excluding the ones, which will be derived from the Sloan Digital Sky Survey (SDSS)
(\cite{Kim}; \cite{Kep}),   
for the Northern sky no automatically extracted catalogs of  
galaxies (nor of clusters) could be produced until   
the recent completion of the Digital Palomar Sky Survey (DPOSS).\\  
DPOSS, which covers three bands (J,F,N), is characterized by   
deeper limiting magnitudes than other catalogs extracted from   
photographic surveys and therefore  
offers an important opportunity to   
investigate galaxy clusters in the nearby (i.e. $z < 0.4$) Universe.   
Recently, DPOSS photometric calibration   
(\cite{Weir95a}) and  extraction of the Palomar Norris Sky Catalog  
(PNSC) were completed at Caltech in collaboration   
with the observatories of Rome, Naples and Rio de Janeiro, partners in the  
CRoNaRio project (\cite{Djor98a}; \cite{Andre97}). This catalog contains astrometric,  
photometric  
and morphological information for all objects detected   
down to limiting magnitudes of $g_J \sim 21.5$, $r_F \sim 20.5$ and  
$i_N \sim 20.0$ in the Gunn \& Thuan photometric system.\\  
With the availability of these new data, various methods to build  galaxy   
cluster catalogs based on  color information have been  
proposed (\cite{Gal99}; \cite{Gal00a}).  
  
   \begin{figure}  
   \resizebox{\hsize}{!}{\includegraphics{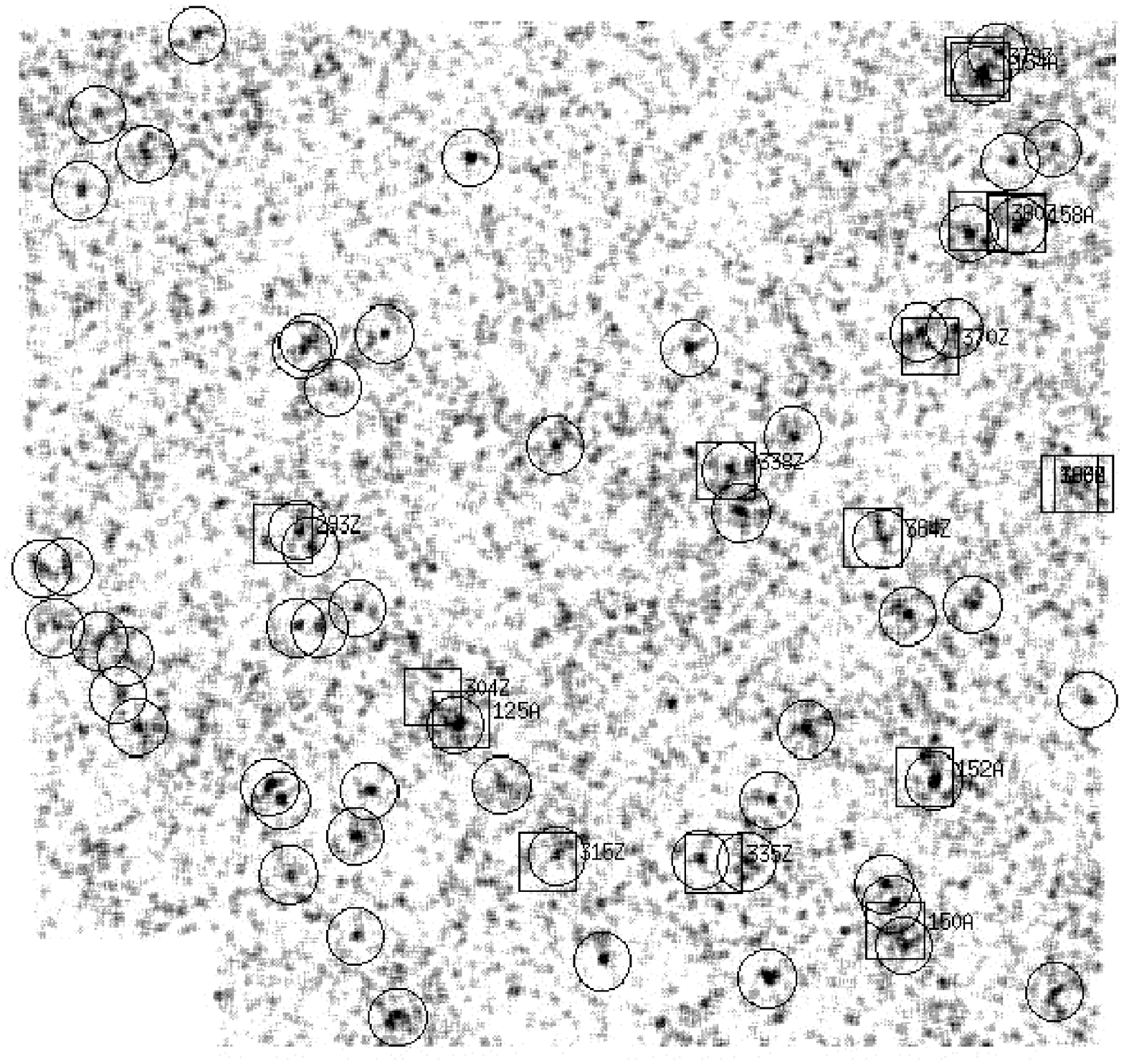}}  
      \caption[]{Detected overdensities in the density map in a    
$5^{\circ}$ x $5^{\circ}$ sky region centered at   
$RA = 1h$ and $Dec=+15^{\circ}$.   
The smoothing on this map is performed by a 3 x 3 Gaussian filter.\\  
Squares mark known Abell and Zwicky clusters. Circles mark putative -  
previously unknown - clusters.}  
         \label{n610}    \end{figure}  
  
In order to exploit the scientific potential of the DPOSS, we developed a model  
independent procedure to search for cluster candidates. The procedure is detailed   
elsewhere and therefore we briefly summarize its main characteristics here   
(\cite{memsait}; Puddu et al., in preparation).  
First, for a given region of the sky, we extract the individual catalogs   
obtained from the corresponding J, F and N DPOSS plates and calibrate them   
to the {\it g, r} and {\it i} Gunn-Thuan System using the procedure described   
in \cite{Weir95b}. Then, after correcting   
for misclassified objects in at least one of the three filters (see \cite{memsait}),   
we produce a   
matched (in the three bands) catalog complete to {\it r} $\sim 19.75$ and   
{\it g} $\sim 20.2$ and, in order to ensure completeness, we disregard all objects  
 fainter then {\it r} $= 19.5$.     
The spatial galaxy distribution in the matched catalog is then binned into   
equal area square bins of $1.2$ $arcmin$ to produce a density map.  
S-Extractor (\cite{SEx}) is run on the resulting image in order to identify   
and extract all the overdensities having number density $2$ $\sigma$ above  
the   
mean background and covering a minimum detection area of $4$ pixels (equivalent   
to $4.8$ $arcmin^{2}$) on the map convolved with a 3 x 3 Gaussian filter. We wish to   
stress that, as in the Schectman (1985)  
approach, we are  not assuming any {\it a priori} cluster model.\\  
The procedure is run on a given plate catalog to  
produce a preliminary candidate cluster catalog.   
All previously known Abell and Zwicky clusters are   
recovered together with many new cluster  
candidates (Fig. \ref{n610} shows, as an example,  DPOSS field 610)   
which need to be confirmed.\\  
The best way to validate cluster finding algorithms would be to apply them to a    
field for which redshifts are available for a fairly deep magnitude limited   
sample of galaxies. Since such a sample does not exist, we were obliged to follow an   
alternative route.   
It is well known that early-type galaxies are preferentially located in the  
cores of rich clusters and that they present a small scatter in the  
color-magnitude diagram (\cite{Dre80}; \cite{DG92}; \cite{Stan98}).   
These properties  
turn the color-magnitude diagram into a powerful tool to disentangle true  
clusters from overdensities of galaxies caused by random object alignment   
along the line of sight.\\   
In this paper we use the early type sequences detected in the  
color-magnitude diagrams obtained from multiband  
optical photometry to confirm a sample of 12 candidates.
These candidate clusters were selected from some fully reduced DPOSS
plates available at the time the test was performed. The selection was
performed randomly in order to include overdensities covering a wide 
range of S/N ratios in the detection maps.
We also use a sample of 8 X-ray selected clusters  
at nearby and intermediate redshift as templates  
to calibrate the photometric redshift estimate.\\

The paper is structured as follows: in section 2 we describe   
the data and the data reduction strategy, in section  
3 we show the color-magnitude diagrams for the calibration  
sample and illustrate the procedure implemented to derive   
the photometric redshift   
estimate, while in section 4 we apply this method  
to the candidates. In section 5 conclusions and future  
developments are discussed.  
  
  
   \begin{table*}[]  
   \caption[]{Selected overdensities.\\  
Notes:(1) Zwicky cluster}  
   \begin{tabular}{ccccccc}  
 \hline  
{OBJ} & {RA(2000)} & {Dec(2000)} & {Richness within} & {Detection} \\  
 &  &  & {the isopleth} & {S/N} \\  
\hline  
27\_694$^{(1)}$& 05 00 07.24 & +10 15 52.00 &  76 & 8 \\  
44\_778 & 08 59 52.68 & +04 10 53.20 &  21 & 4 \\  
17\_778 & 09 08 28.20 & +06 03 39.55 &  63 & 8 \\  
5\_778  & 09 12 11.09 & +02 23 10.22 &  17 & 4 \\  
1\_778  & 09 12 15.34 & +02 32 18.11 &  48 & 7 \\  
64\_781 & 09 57 25.10 & +03 39 06.70 &  27 & 5 \\  
72\_781 & 09 57 53.23 & +03 27 10.09 &  59 & 7 \\  
6\_725  & 15 24 52.60 & +11 20 27.10 & 146 & 12 \\  
1\_799$^{(1)}$& 16 03 11.78 & +03 14 17.63 & 132 & 11 \\  
24\_694 & 05 03 38.92 & +10 38 08.59 &  50 & 7 \\  
21\_694 & 05 04 42.34 & +10 48 49.00 &  70 & 8 \\  
26\_727 & 15 57 48.70 & +08 52 04.39 &   8 & 3 \\  
\hline  
\end{tabular}  
   \label{OVtab}  
\end{table*}  
  
  
   \begin{table*}[]  
   \caption[]{Sample of known clusters.\\   
G\&L: Gioia \& Luppino, 1994; S\&R: Struble \& Rood, 1999.}  
   \label{MStab}  
   \begin{tabular}{ccccccc}  
 \hline  
{Id} & {RA(J2000)} & {Dec(J2000)} & {z} & {Ref.(z)} \\  
  
\hline  
MS0821.5+0337 & 08 24 07.104 & +03 27 45.22 & 0.347 & G\&L\\  
Abell 1437 & 12 00 24.960 & +03 20 56.40 & 0.1339 & S\&R\\  
MS1253.9+0456 & 12 56 28.827 & +04 40 01.87 & 0.23 & G\&L\\  
Abell 1835 & 14 01 02.399 & +02 52 55.20 & 0.2532 & S\&R\\  
MS1401.9+0437 & 14 04 29.378 & +04 23 00.33 & 0.23 & G\&L\\  
MS1426.4+0158 & 14 28 58.768 & +01 45 11.94 & 0.32 & G\&L\\  
Abell 2033 & 15 11 23.518 & +06 19 08.40 & 0.0818 & S\&R\\  
MS1532.5+0130 & 15 35 02.739 & +01 20 57.15 & 0.49 & G\&L\\  
\hline  
\end{tabular}  
\end{table*}  
  
\section{Observations and Data Reduction}  
  
\subsection{The observations and the sample}  
  
All data used in this paper were obtained in imaging mode with   
DFOSC at the ESO 1.54m Danish telescope (La Silla - Chile) during two   
observing runs (March 1999 and March 2000) blessed by dark time and  
photometric conditions.   
The CCD (LORAL/LESSER C1 W7) has 2052 x 2052 pixels, each pixel covering   
$0.39^{\prime\prime}$, corresponding to a field of $\sim$ $13.3^{\prime}$   
x $13.3^{\prime}$.    
Data were taken in the {\it g}, {\it r} and {\it i} filters of  
the Thuan \& Gunn system (\cite{TG76}; \cite{Wade}).  
Seeing averaged $\sim 1.2^{\prime\prime}$ and was always better than  
$1.5^{\prime\prime}$.  
The slight difference between our setup and the original Thuan \& Gunn   
filters resulted in a significant color correction which had to be taken   
into account in the calibration procedure.\\   
Exposure times ranged from 40 to 50 minutes in the {\it g} band  
and 20 to 30 minutes in the {\it r}  and {\it i} bands, depending on the target.   
In order to obtain higher S/N and  more accurate photometric  
measurements, exposures for those clusters at intermediate redshift   
were usually repeated in two or three slightly offset frames.\\  
The observed fields (all located in a region with $0^{\circ} < \delta <  
+12^{\circ}$ and therefore observable from both hemispheres) included   
12 candidate clusters plus 8 clusters with known  
redshifts to be used as comparison sample.   
In order to use the same material to both calibrate the corresponding DPOSS   
fields and to validate our algorithm, we selected galaxy overdensities in   
such a way that we had at least two (up to four) candidates and/or  
clusters in each DPOSS field. One candidate, observed on two different  
nights, also provided an independent check of the photometric accuracy of the  
second run.   
For the comparison sample  we selected 8 clusters from  
the X-ray selected sample of \cite{GL94} and \cite{Ebe96}.\\   
The equatorial coordinates  of the observed fields are given  
in Tables \ref{OVtab} and Tables \ref{MStab}.\\  
We wish to stress that one of the main problems encountered in our work was   
the well known lack of a suitable set of photometric standards for the   
Gunn-Thuan system which, along with the lack of faint stars suitable for CCD   
observations, very often prevents  good coverage of the airmass-color plane.   
The problem is even worse for observers in the Southern hemisphere where   
the number of available standards is uncomfortably small.   
We succeeded, however, in observing an average of 4-5 standard stars per observing   
night.   
  
\subsection{Data Reduction and Photometric Calibration}  
  
The raw images (both scientific and calibration) were prereduced using the   
standard procedures available in the IRAF package.   
First, the frames were corrected for instrumental effects (overscan   
and bias) and flat fielded. Individual dome and sky flats in each filter  
were  median stacked to increase the S/N ratio.   
For the first run, flatfielding was performed using sky flats only, but the   
experience gained in this run suggested  a slightly different   
procedure for the second run, using dome flats to achieve  better  
correction of the small scale pixel-to-pixel variations.  
Dome flats were first used to correct the sky flat frames for the higher  
frequency fluctuations, and the resulting frames were then smoothed and stacked   
to map the lower frequency fluctuations and combined with the average   
dome flat to produce the final master flats.\\  
In the first run, we divided the exposures into two or three frames for the same field; in  
these   
cases, the images were combined in each filter by medianing (three exposures)   
or averaging (two exposures) the aligned frames.  
Standard star photometry was performed using the {\it apphot} package in IRAF.   
Due to the need to defocus most of the stars to avoid saturation, stars   
were measured through 10 apertures with diameters up to $90$ pixels  
($35.1^{\prime\prime}$), and the local sky was determined using a $10$   
pixel wide annulus outside of the largest aperture.  
In order to determine the zero-point offset and the airmass and color terms   
we used the $40$ pixels ($\simeq 15^{\prime\prime}$) aperture, for the  
focused and unsaturated stars, and the asymptotic magnitudes for the   
defocused ones.\\  
  
\begin{figure}  
   \resizebox{\hsize}{!}{\includegraphics{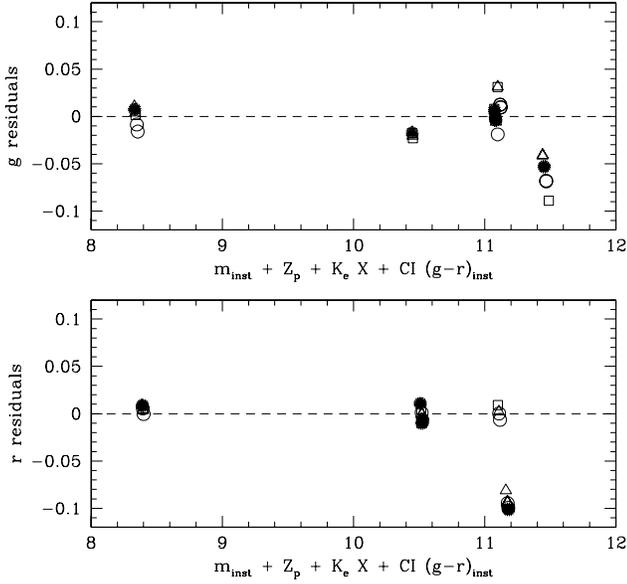}}  
      \caption[]{Residuals for the {\it g} (upper panel)  
and {\it r} (lower panel) fit. Different symbols refer to different nights.  
The outlier point is Ross 683, which turns out to be   
consistently brighter than expected in all four observed nights.}  
         \label{cal}     
    \end{figure}

The IRAF {\it fitparams} task was used to fit  the data with the relation  
\begin{equation}  
{m_{true} = m_{inst} + Z_{p} + K_{e} \cdot X + CI \cdot color_{inst}}  
\end{equation}  
where $Z_{p}$ is the zero point of the magnitude scale; $K_{e}$ is the   
extinction coefficient and $X$ the airmass; $CI$ is the instrumental color   
term coefficient.   
For the first run (March 1999), due to the paucity of standard stars,  
we could determine single night coefficients only for the {\it r} and  
{\it i} filters.  
The coefficients were consistent from one night to the other  
and we used a mean fit for the {\it g-r} color. \\  
For the second run (March 2000) we  instead derived the coefficients for  
each night and in each band (using the {\it g-r} and {\it r-i} colors).   
The resulting calibration coefficients of the various nights are   
consistent within the errors and, therefore, in order to improve the quality   
of the fit, we adopted a unique pair of extinction coefficient and color term   
for the whole run. These constants were then used to derive the zero  
points for each night.   
In Fig. \ref{cal} we show for the {\it g} and {\it r} filters,   
the fit residuals as a function of the estimated magnitude,  
using different symbols for different nights.\\  
  
   \begin{figure}  
   \resizebox{\hsize}{!}{\includegraphics{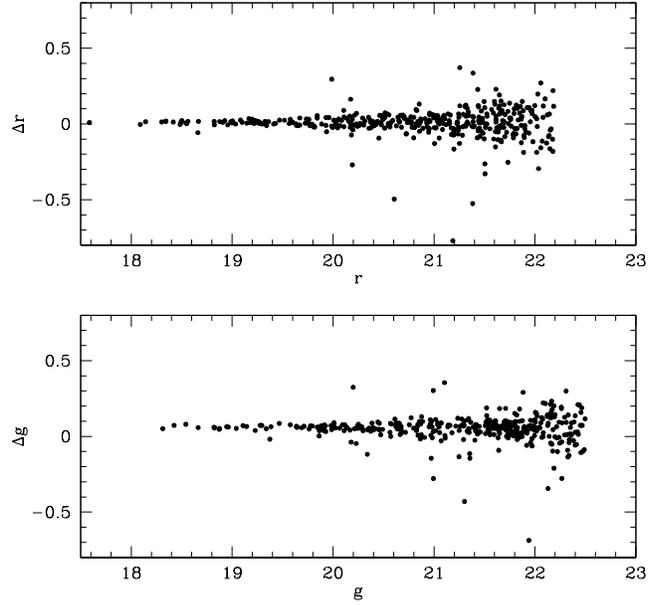}}  
      \caption[]{Comparison between the photometry of the candidate  
cluster 64\_781 between nights 2 and 4: the plot shows the {\it  
g} or {\it r} magnitude offset vs. the {\it g} or {\it r} magnitude}  
         \label{n2n4}      
   \end{figure}  
  
   \begin{figure}  
   \resizebox{\hsize}{!}{\includegraphics{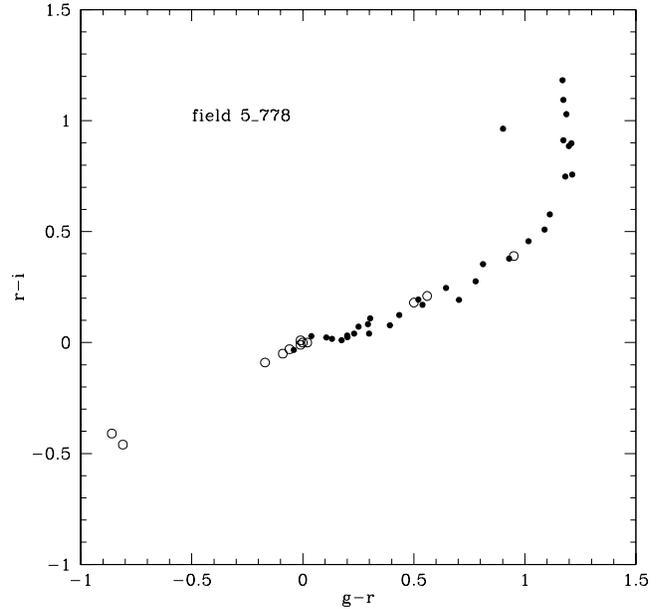}}  
      \caption[]{Color - color diagram for the Gunn \& Thuan standards  
(open circles) and for a sample of stars (filled symbols) extracted from the field  
5\_778.}  
         \label{cc5778}      
   \end{figure}  
  
\subsection{Object Detection and Photometry}  
  
The object catalogs were produced individually for each  
band using S-Extractor: all  objects larger than $4$  
pixels and $2\sigma$ above the background counts were included  
and their photometric and morphological features measured.   
We used a photometric reference aperture with a diameter $\sim3$ times larger than the average seeing.\\  
For each CCD field, the three single band catalogs where  
matched taking into account  
the  shifts between pointings (measured using  
the {\it geomap} and {\it geotrans} IRAF tasks).\\  
  
To obtain an estimate of the external photometric errors, the   
candidate cluster  64\_781 was observed on two  
different nights. In this way we could evaluate  possible  
night-to-night magnitude offsets in both the {\it g} and {\it r} filters. The  
typical weighted mean values for these offsets are $0.05$ for the  
{\it g} filter and $0.007$ for the {\it r} filter, i.e. they are of   
the same order as the rms errors from the three parameter calibration fit  
(Fig. \ref{n2n4}).\\  
  
Since our goals require  high accuracy for the  
color determination, we  further checked   
the photometric calibration, using the following test: in the color-color diagram   
(Fig. \ref{cc5778}) we plotted the linear sequence of the  
Gunn-Thuan standards (open circles) together with all  
the unsaturated stars (S-Extractor stellarity index $< 0.8$)   
 within the limiting magnitude, selected from  
some CCD cluster candidate fields (in  Fig. \ref{cc5778} we show   
 the 5\_778 field). For all of these fields,  
the sequence of the selected stars is linear (excluding the  
very red sources, dominated by stars of spectral type M,  
which have a constant {\it g-r} color while {\it r-i} depends  
on the spectral subtype; see \cite{Fuku96}  
and \cite{Fin00}) and overlap quite well with the standard sequence.   
This means that the colors of the main  
sequence stars are well determined, since the relation  
between {\it g-r} and {\it r-i} is the same for the cluster field stars  
and for the standards.  
  
\section{Validation method via  color-magnitude diagrams}  
  
From the {\it g} and {\it r} matched catalogs  
we excluded obvious stars (stellarity index  
$> 0.95$) and then  selected a box of $\sim 300$ pixels in size 
(corresponding to a typical 
core cluster diameter of $\sim 750$ kpc at $z \sim 0.2$),  
centered on the approximate cluster center and a second  
box of equal size located as far as possible from the cluster,  
to be used for the evaluation of the background contribution.   

Our procedure is summarized in Fig. \ref{grall}, which shows the results  
 for four sample candidates which are  representative of the  
various morphologies encountered.\\  
For all candidates in our sample we first obtained the color-magnitude   
diagrams for both the cluster and background objects. Then, in order to   
enhance the early-type sequence, we performed the statistical subtraction of the  
background contribution by eliminating  
for each object in the background diagram the corresponding nearest  
galaxy in the cluster+background diagram.   
If we then isolate the objects contained within a narrow strip of  
the color-magnitude diagram centered around the early-type sequence,   
the galaxy overdensities become more evident in both the spatial   
distribution and in the number counts radial profile  
(Fig. \ref{grall}; the radial profile was calculated by choosing as   
cluster center the barycenter of the density distribution).\\  
The plots in Fig. \ref{grall} can  
be  used as a criterion to distinguish true clusters (candidates  
(a) and (b)), even if they are difficult to detect.    
In some cases (usually  candidates which are either  
too distant or too poor), despite the presence of an apparent sequence 
in the color-magnitude diagram, 
the objects do not form a physical overdensity, but  
turn out to be uniformly distributed on the sky. In these cases  
it is more difficult to reach any definite conclusion about the physical  
nature of the candidate.  
  
\subsection{Color-magnitude diagrams for the calibration  
cluster sample}  
  
We chose as templates a sample of X-ray clusters for which,  
at least in principle, the early-type sequence in the  
color-magnitude diagrams should be easily detectable.  
This sample was also used   
to investigate whether or not it was possible to derive an acceptable  
estimate of the redshift from the {\it g-r} color of the early-type  
sequences.\\  
  
\begin{figure*}  
   \resizebox{\hsize}{!}{\includegraphics{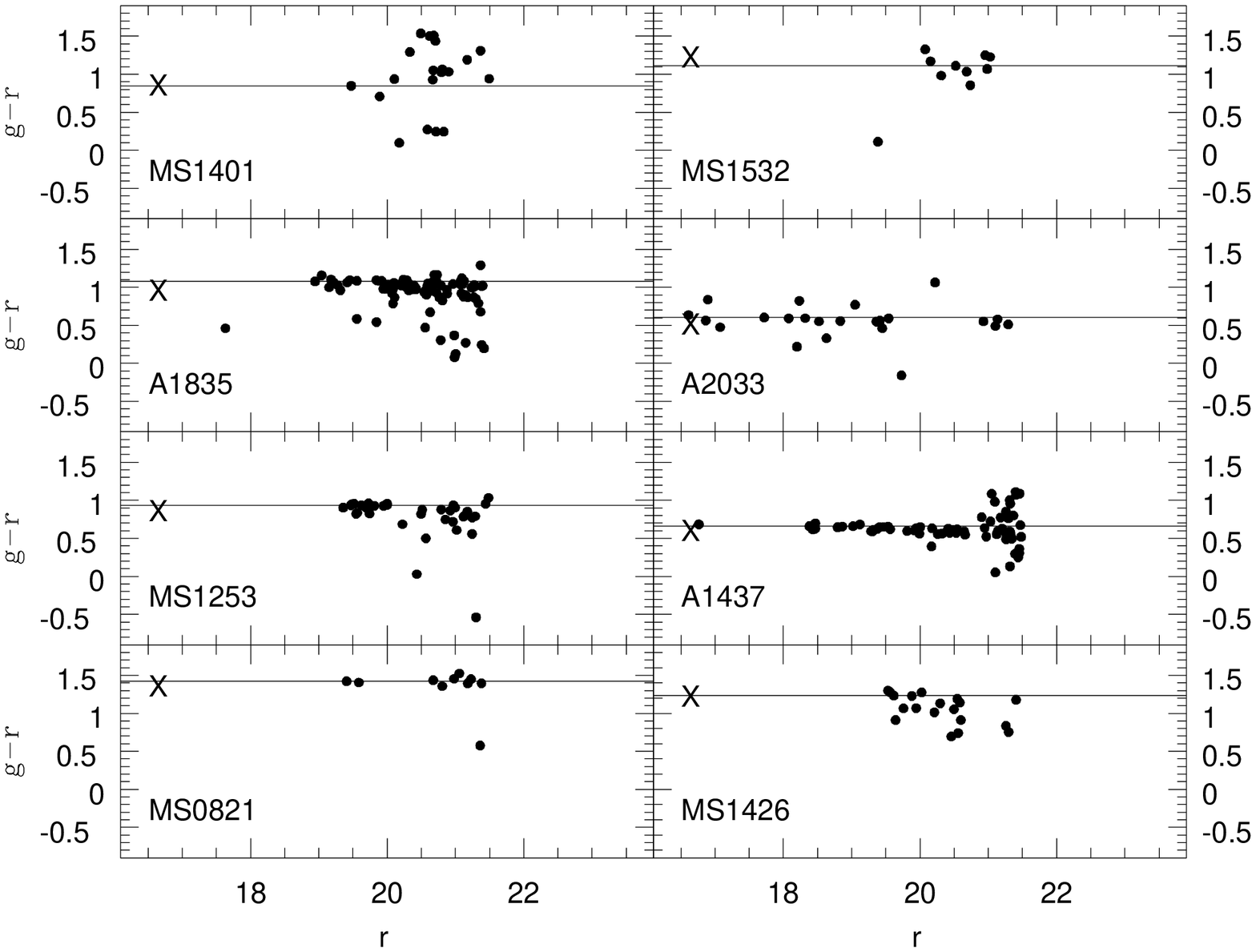}}  
      \caption[]{Color-magnitude diagrams after statistical background  
subtraction for the X-ray cluster sample. According to \cite{GL94},  
MS1401 is a loose cluster, without a dominant galaxy; MS1426 contains    
spirals and possibly interacting systems, and moreover may be two clusters   
in projection. A Seyfert galaxy at z=0.074 is present in the foreground  
of the MS1532 field. Abell 2033 is of Bautz-Morgan type III and richness class 0.}  
         \label{grnoti}      
   \end{figure*}

In Fig. \ref{grnoti} we plot the color-magnitude diagrams for the 8 clusters  
in the X-ray sample including only the cluster contribution (i.e., after the  
statistical subtraction of the background).  
It is quite evident that some of the early-type sequences are  
only broadly outlined (MS1401, MS1426, MS1532), which may  
 be caused either by the intrinsic faintness of  
the cluster members or by  cluster structural features  
(poorness, looseness and presence of interacting systems;  
see the comments in \cite{GL94} about these  
three clusters). For each cluster we derived a median {\it g-r} color  
using only the  5 brightest galaxies after the background  
subtraction (continuous line),   
from which we also estimated the redshift.  
The crosses represent the {\it g-r} colors corresponding to the  
literature redshifts.  
  
\subsection{Photometric Redshift Estimate}  
  
Some techniques for deriving redshifts from broadband photometry   
consist of matching  observed elliptical galaxy  
colors with those predicted from the Spectral Energy  
Distributions (SEDs) (\cite{VS77}) of  
a template elliptical galaxy at zero-redshift and corrected  
according to the redshift: since ellipticals become redder as  
their redshift increases and since the redshift dependent   
correction (k-correction), is monotonically increasing in the near  
and intermediate redshift Universe, colors can be used to infer  
the cluster redshift.\\   
There is no agreement in the literature for the {\it g-r} color  
of ellipticals at zero-redshift: $0.47^m$, according   
to \cite{SGH83}; $0.38$ mag for \cite{FG94}; $0.40^m$ to $0.49^m$  
according to \cite{Fuku95}. Differences in  
these values likely depend on the galaxy spectrum template   
adopted for the ellipticals and on the use  
of a synthetic or an observed spectrum for the standard  
stars defining the photometric system. To a lesser extent,  
differences are due to the variations in the actual shape  
of the Gunn {\it g} and {\it r} filters (convolved with the   
atmosphere, mirror and glass transmissions, CCD quantum  
efficiency, etc.) and possibly also to the  
way in which the colors are computed.  
  
   \begin{figure*}  
  
    \resizebox{!}{9cm}{\includegraphics{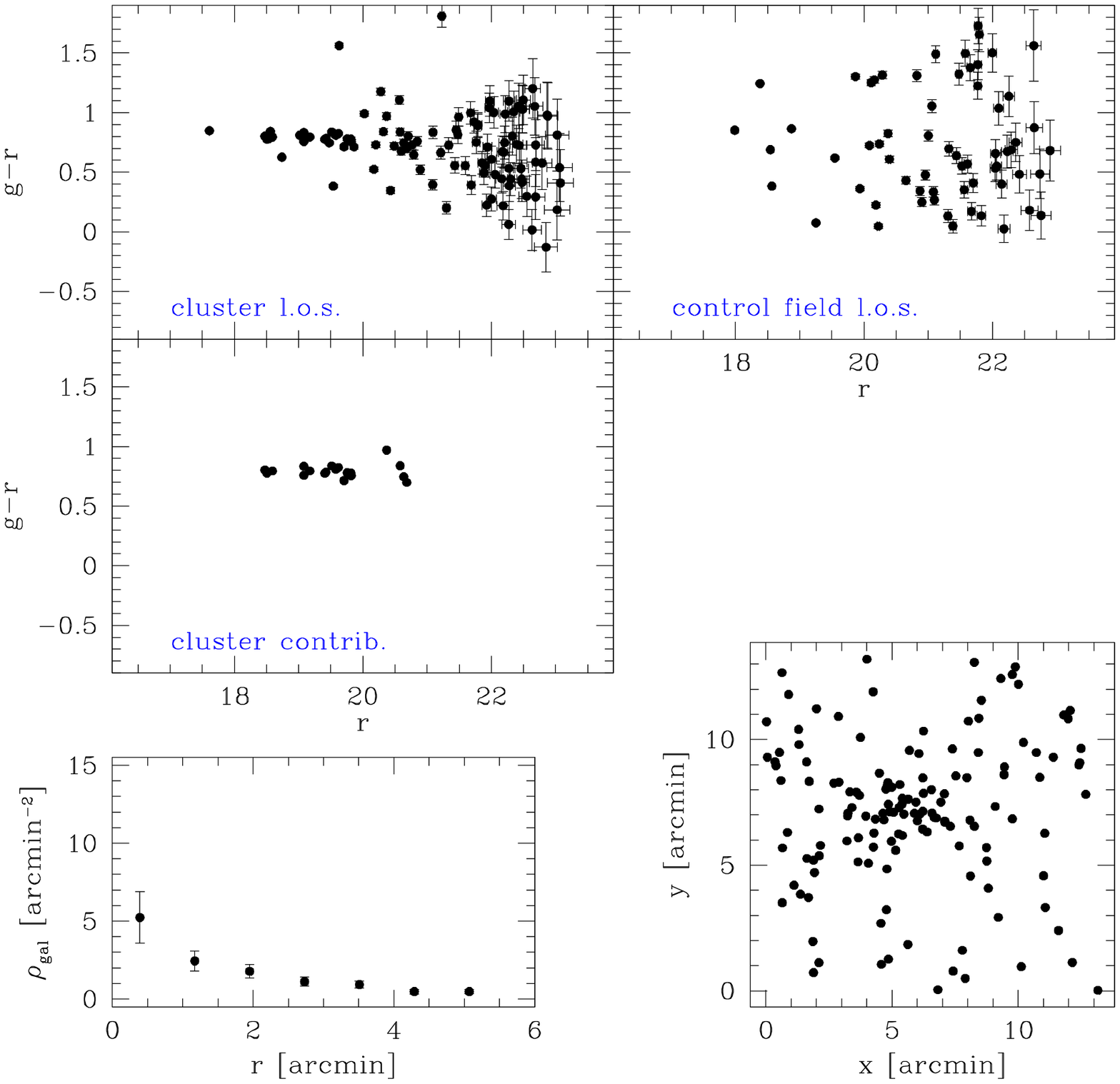}}  
    \resizebox{!}{9cm}{\includegraphics{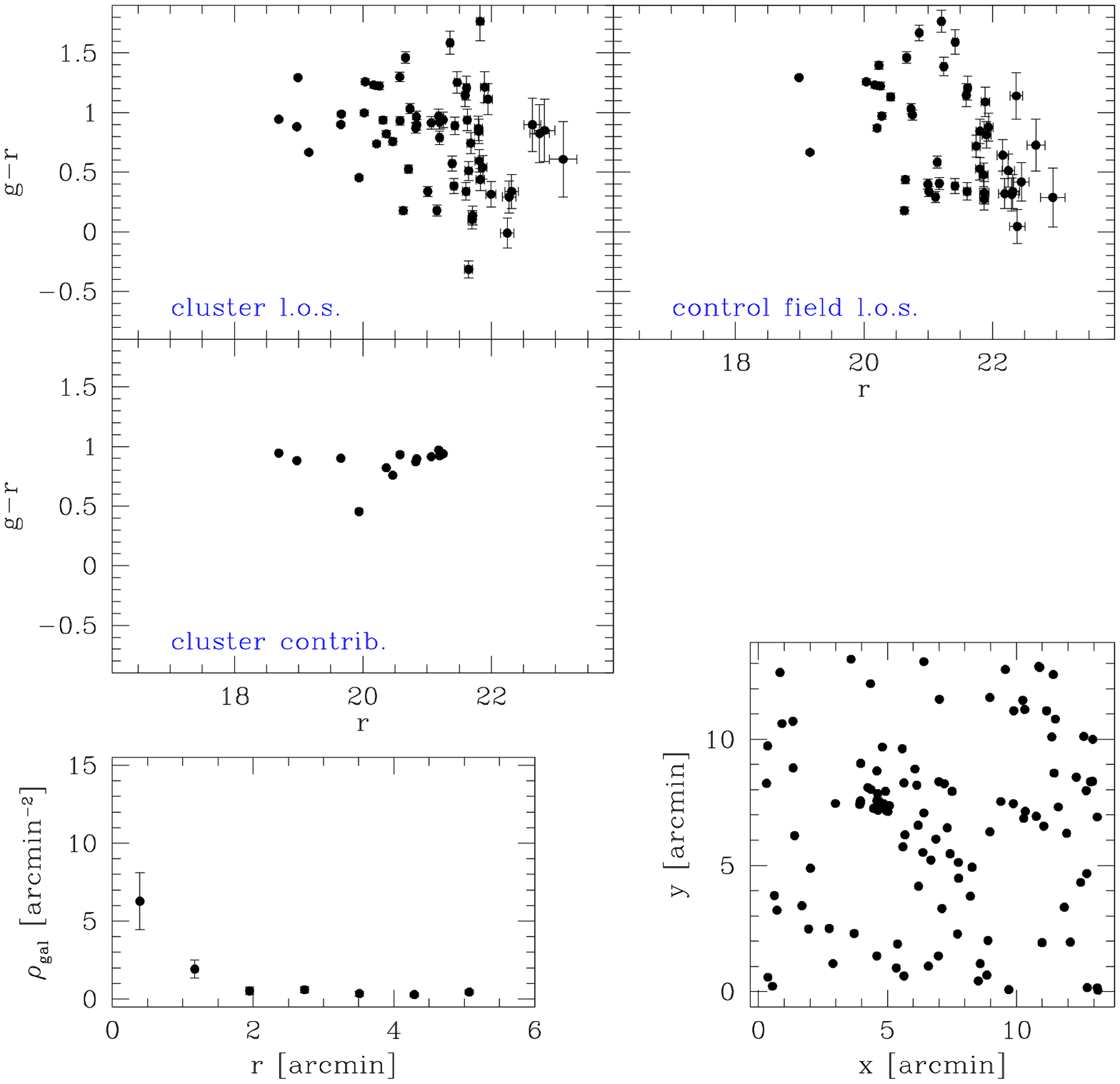}}  
	\begin{center} (a) \hspace{8cm} (b) \end{center}   
    \resizebox{9cm}{!}{\includegraphics{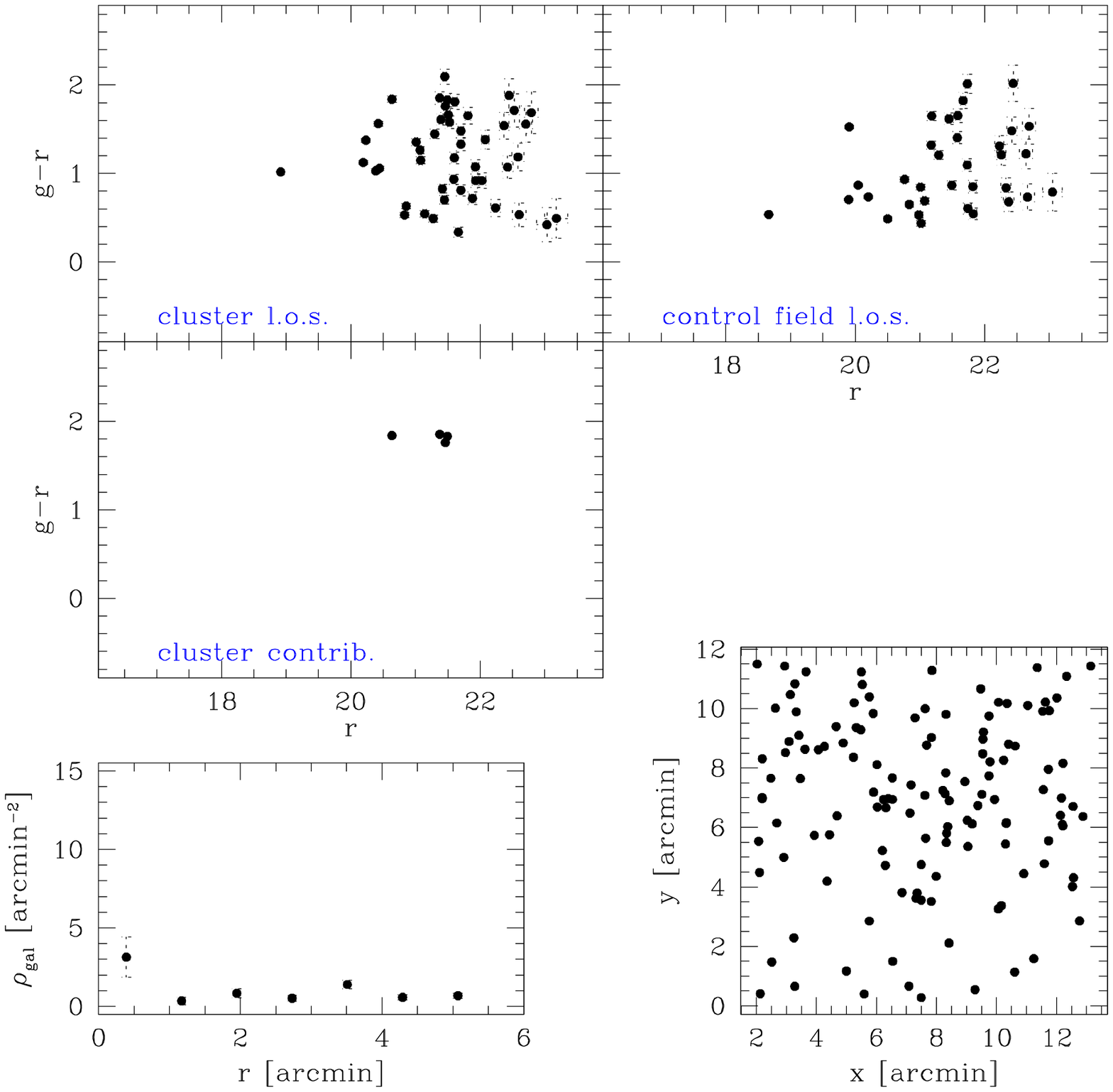}}  
    \resizebox{9cm}{!}{\includegraphics{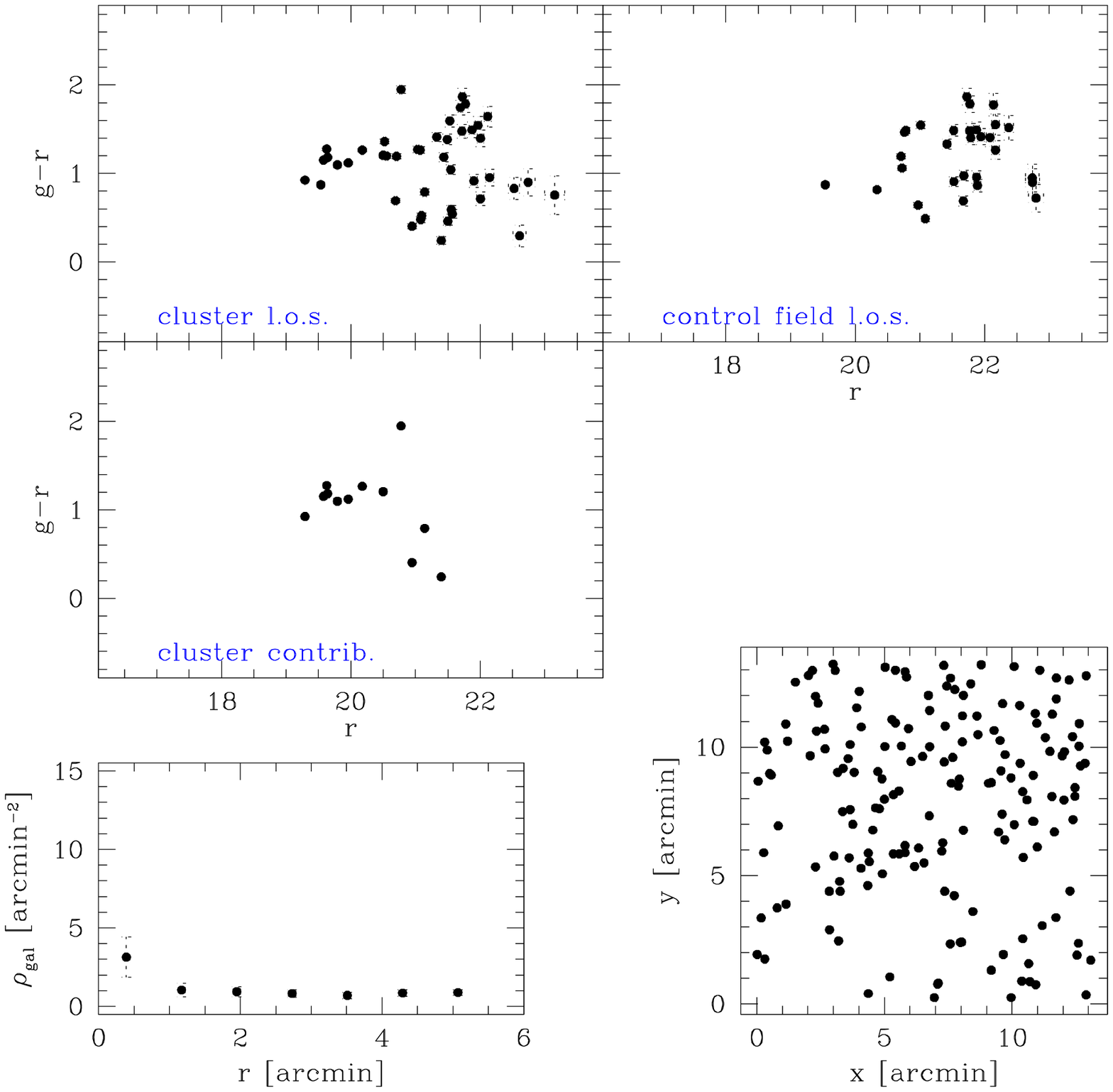}}  
	\begin{center} (c) \hspace{8cm} (d) \end{center}   
  
      \caption[]{Complete set of plots for two confirmed (upper half) and  
two uncertain (lower half) cluster candidates. Each set of plots consists  
of color-magnitude diagrams for background + cluster box; color-magnitude diagrams  
for background box; color-magnitude diagrams for statistically corrected cluster   
objects; radial profiles and spatial distributions over the CCD field  
of the objects within a color strip around the early-type sequence.  
The four sets of plots refer respectively to: (a) OV17\_778, a typical rich  
nearby cluster (z$\sim$ 0.2); (b) OV6\_725, a poor cluster at redshift  
z$\sim$ 0.2; (c) OV21\_694 and (d) OV24\_694, uncertain clusters,   
with less evidence for the early-type sequence in the color-magnitude  
diagrams, and for density peaks in the spatial distribution.}  
         \label{grall}      
   \end{figure*}  
  
\begin{figure}  
   \resizebox{8cm}{8cm}{\includegraphics{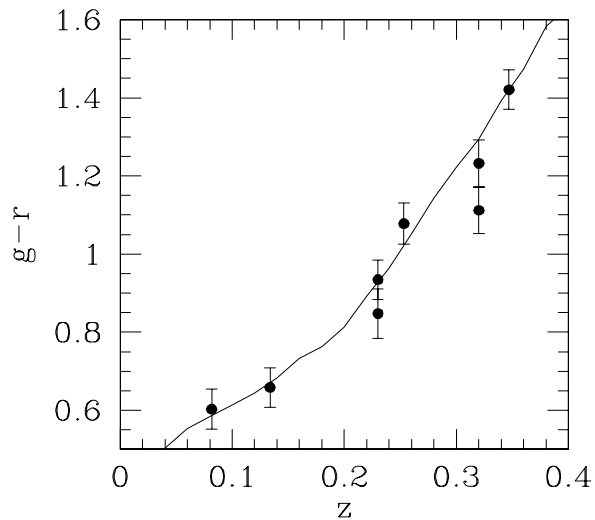}}  
      \caption[]{The observed colors from the color-magnitude relation and  
their errors for the X-ray cluster sample (filled circles)  
are plotted, compared to the expected color of ellipticals (continuous line)   
as a function of spectroscopic redshift.}  
        \label{sgh}      
   \end{figure}

\begin{figure}  
   \resizebox{8cm}{8cm}{\includegraphics{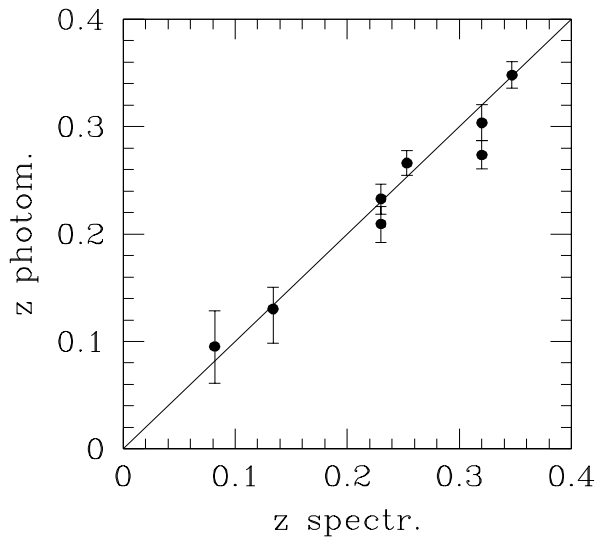}}  
      \caption[]{Photometric vs. spectroscopic redshift. The solid line  
is the bisector and is not derived from data fitting.}  
        \label{zz}      
   \end{figure}

In the absence of a definite value, we left the zero-redshift  
color of ellipticals as the unique free parameter and constrained   
it with our own observations by (robustly) fitting the   
relation between color and redshift. Fig. \ref{sgh}  
shows (filled dots) the observed colors from the color-magnitude 
relation 
vs. known spectroscopic redshift for our X-ray cluster sample. 
The expected color 
of ellipticals (continous line) were computed using  
the \cite{SGH83} k-correction curve and our  
own determination of the elliptical colors. The average {\it g-r}
color of ellipticals of zero redshift turns out to be $0.44^m$, i.e. the 
average of the four previously quoted literature values. Errors on  
the colors are given as one third of the interquartile range,  
which roughly corresponds, for a Gaussian distribution of five points   
to the error on the mean. We prefer these to the standard  
error since they are more robustly determined. Figure \ref{sgh}  
shows that all points are compatible with the curve within $1\sigma$,   
excluding two points , which are
within $2\sigma$. The agreement is good, provided that there is  
only one free parameter (the rest-frame elliptical color).  
Figure \ref{zz} compares the photometric redshift, estimated  
from the color-magnitude diagrams and the spectroscopic  
redshift. The agreement is good, and the error (interquartile range)   
on the redshift is, on average, $\delta z = 0.01$, i.e. $3000$ km/s.  
Tab. \ref{zztab} lists the estimated photometric redshifts, with the errors computed as  
previously defined, for the putative clusters;  
since these clusters are fairly rich systems, this error is likely to be a lower 
limit.

\section{The validation of the candidate cluster sample}  
  
\begin{figure*}  
   \resizebox{\hsize}{!}{\includegraphics{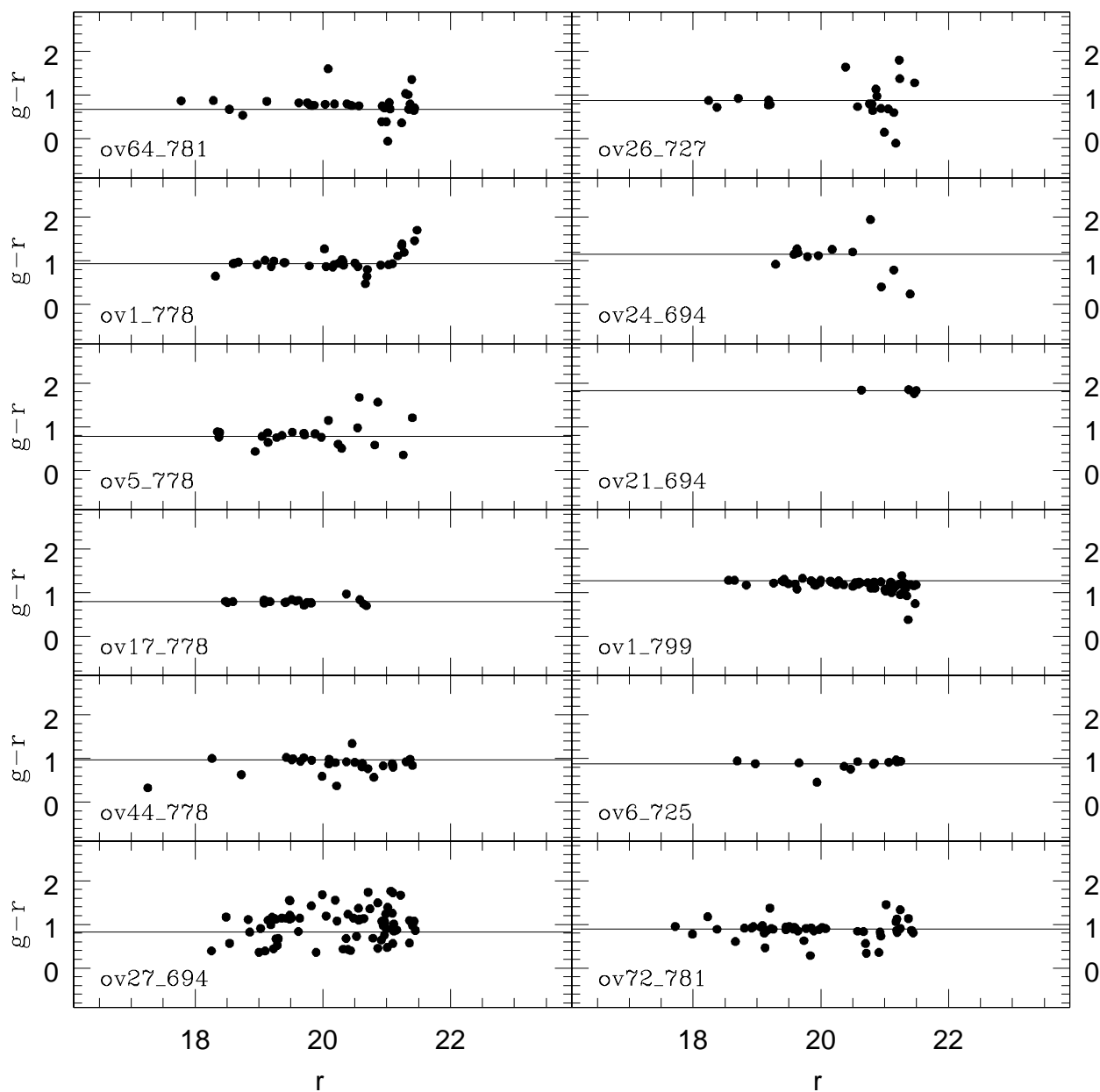}}  
      \caption[]{Color-magnitude diagrams after statistical background  
subtraction for the 12 candidate clusters with complete observations. The  
three upper right diagrams refer to not confirmed candidates (OV21\_694,  
OV24\_694, OV26\_727).}  
         \label{grcand}      
   \end{figure*}

In Fig. \ref{grcand} we show the early-type sequences obtained (after   
subtracting  
the background) for the 12 candidate clusters in our sample. We    
confirm 9 of the 12 candidates as true clusters,
one is a fortuitous detection of a cluster at $z \sim 0.4$,   
one is a false detection and the last is undecidable on the  basis   
of the available data.   

The better definition of the early-type sequences observed in the DPOSS  
confirmed clusters sample with respect to the X-ray sample is likely   
due to the different specific properties of the two samples:
at a given z, our optically selected clusters are on the average   
richer and more centrally concentrated then the
X-ray selected ones.

For the cluster  OV27\_694, the early-type sequence is less defined due to  
the overlap of two independent clusters/groups along the  
line of sight.\\  
The case of OV21\_694 merits special attention.  
 As a visual inspection of the corresponding POSS-II F plate  
shows, the marginal detection of an early type sequence (Fig. \ref{grall}   
and \ref{grcand})  
seems to be due to the chance alignment of a distant cluster (at  
redshift  $z \sim 0.4$) with a rich galaxy field. The existence of such a  
foreground rich field has therefore triggered the search algorithm.  
As far as OV24\_694 and OV26\_727 are concerned, the marginal evidence for   
an early-type sequence does not correspond to a defined  
overdensity in the number count radial profiles (using galaxies in the  
 strip centered on the mean g-r color).  
The case of OV26\_727 is a false cluster detection, since it has a low S/N ratio  
and low isophotal richness (see Tab. \ref{OVtab}; it may be a group).  
Visual inspection of the POSS-II plate shows that OV24\_694 lies in  
a crowded field rich with galaxies; from the sky diagram (Fig. \ref{grall},   
case (d)) it is also evident that a large fraction of these foreground galaxies   
have the same color. Thus, this field could be  part of a larger loose cluster   
or a cluster in a region with variable background.

   \begin{table}[htb]  
   \caption[]{New Redshift Estimate.\\  
(1): the faintness  
of the galaxies, as observed both on CCD and on plate    
suggests that the candidate cluster is far, as puts forward by  
the redness of the color-magnitude relation;   
(2): the color-magnitude shows a large scatter,   
suggesting that this cluster is possibly contaminated by a   
foreground group.}  
   \begin{tabular}{cccc}  
\hline  
{OBJ} & {z est.} & {z est. min} & {z est. max}\\  
\hline  
17\_778 & 0.195 & 0.169 & 0.209 \\  
1\_778 & 0.234 & 0.219 & 0.247 \\  
1\_799 & 0.314 & 0.299 & 0.326 \\  
21\_694$^{(1)}$ & 0.488 & 0.466 & 0.511 \\  
24\_694 & 0.282 & 0.271 & 0.295 \\  
26\_727 & 0.216 & 0.204 & 0.23 \\  
27\_694$^{(2)}$ & 0.204 & 0.183 & 0.218 \\  
44\_778 & 0.243 & 0.228 & 0.255 \\  
5\_778 & 0.189 & 0.159 & 0.205  \\  
64\_781 & 0.139 & 0.110 & 0.159 \\  
6\_725 & 0.218 & 0.205 & 0.232  \\  
72\_781 & 0.219 & 0.207 & 0.234 \\  
\hline  
        \end{tabular}  
   \label{zztab}  
\end{table}  
  
\section{Summary and conclusions}  
  
The aim of our work was to test the validity of a model independent cluster  
finding algorithm, implemented to extract a statistically well defined sample   
of cluster candidates from photometrically  
calibrated DPOSS data (see Paper I for details).  
The advantages of a model independent approach are that i) the program  
does not assume any a priori knowledge about the clusters, and ii) it  
objectively looks for statistically meaningful overdensities in the  
galaxy density field.  
The main problem in validating any cluster finding algorithm is the lack  
of a suitable data set to use as a template, i.e., the lack of a  
region of the sky containing a large sample of clusters with well  
defined redshifts and properties.  
In the absence of such a data set, we adopted a photometric approach based  
on the use  of the sequence defined in the  
color-magnitude diagrams of clusters by bright early-type galaxies as a diagnostic tool.  
We obtained deep multiband CCD photometry for a sample of 12 candidate  
clusters extracted from the DPOSS data, plus an additional sample of 8  
X-ray clusters with known redshifts to be used as a template to  
calibrate the photometric redshift procedure.  
Results may be summarizied as follows: among the 12 clusters candidates,  
10 are confirmed clusters, 1 is false and 1 is uncertain. The X-ray  
selected cluster sample was then used both to check the accuracy  
($\Delta z \simeq 0.01$) and to find the zero point (i.e., the average  
zero-redshift $g-r$ color for elliptical galaxies in our system:$(g-r)_0  
=0.44$) for the photometric redshift procedure.  
This procedure is being applied to a larger sample of clusters derived  
from both DPOSS calibration data and from other archive datasets.  
Future papers will deal with the analysis of a larger sample of clusters  
($\sim 200$) identified on both DPOSS and archive data and will focus  
on the derivation and analysis of luminosity functions (both individual  
and cumulative) and of radial number count profiles (\cite{thver}).  
  
{}


\begin{thebibliography}{}    
  
 \bibitem[Abell 1958]{Abell1} Abell, G.O. 1958, A\&AS, 3, 211  
 \bibitem[Abell et al. 1989]{Abell2} Abell, G.O., Corwin, H.G., Olowin,  
	R.P. 1989, AJSS, 70, 1  
 \bibitem[Andreon et al. 1997]{Andre97} Andreon, S., Zaggia, S., de  
	Carvalho, R., et al. 1997, in {\it Rencontres de  
 	Moriond}, eds. G. Mamon, T. X. Thuan and Y. T. Van,  
 	Edition Frontieres (Gif-sur-Yvette).  
 \bibitem[Bertin \& Arnouts 1996]{SEx} Bertin, E. \& Arnouts, S. 1996,  
	AASS, 117, 393  
 \bibitem[Couch et al. 1991]{Couch91} Couch, W.H., Ellis, R.S., Malin,  
	D.F., et al. 1991, MNRAS, 249, 606  
 \bibitem[Dalton et al. 1992]{Dal92} Dalton, G.B., Efststhiou, G., Maddox,  
	S.J., et al. 1992, ApJ, 390, L1-L4  
 \bibitem[Djorgovski et al. 1998b]{Djor98b} Djorgovski, S.G., de Carvalho,  
	R.R., Gal, R.R., et al. 1998, in {\it IAU Symp. 179},   
	McLean, B.J., Golombeck, D.A., Hayes,  
	J.J.E., Payne, H.E. eds., Kluwer Academic Publ., p. 424  
 \bibitem[Djorgovski et al. 1998a]{Djor98a} Djorgovski, S.G., Gal, R.R.,  
	Odewahn, S.C., et al. 1998, in {\it Wide  
	Field Surveys in Cosmology}, S. Colombi, Y. Mellier, and  
	B. Raban, Gif sur Yvette eds., Eds. Fronti\`eres, p. 89  
 \bibitem[Dodd \& MacGillivray 1986]{DMac86} Dodd, R.J., MacGillivray,  
	H.T. 1986, AJ, 92, 706  
 \bibitem[Dressler 1980]{Dre80} Dressler, A. 1980, ApJ, 236, 351  
 \bibitem[Dressler \& Gunn 1992]{DG92} Dressler, A. \& Gunn, J.E. 1992,   
	ApJS, 78, 1  
 \bibitem[Ebeling et al. (1996)]{Ebe96} Ebeling, H., Voges, W., Bohringer,  
	H., et al. 1996, MNRAS, 283, 1103  
 \bibitem[Finlator et al. 2000]{Fin00} Finlator, K., Ivezic, Z., Fan, X., et  
	al. 2000, AJ, 120, 2615  
 \bibitem[Frei \& Gunn (1994)]{FG94} Frei, Z., Gunn, J.E. 1994, AJ, 108,  
	1476  
 \bibitem[Fukugita et al. (1995)]{Fuku95} Fukugita, M., Shimasaku,  
	K.,Ichikawa, T. 1995 PASP, 107, 945  
 \bibitem[Fukugita et al. 1996]{Fuku96} Fukugita, M., Ichikawa, T., Gunn,  
	J.E., et al. 1996 AJ, 111, 1748  
 \bibitem[Gal et al. 1998]{Gal98} Gal, R.R., de Carvalho, R.R.,  
	Djorgovski, S.G., et al. 1998,  
	{\it American Astronomical Society Meeting}, 193, 202  
 \bibitem[Gal et al. 1999]{Gal99} Gal, R.R., Odewahn, S.C., Djorgovski,  
	S.G., et al. 1999,  
	in {\it Photometric Redshifts and Detection of High Redshift  
	Galaxies}, ASP Conference Series, vol. 191, Eds. Ray  
	Weymann et al., p. 185  
 \bibitem[Gal et al. 2000a]{Gal00a} Gal, R.R., de Carvalho, R.R., Odewahn,  
	S.C., et al. 2000 AJ, 119, 12  
 \bibitem[Gal et al. 2000b]{Gal00b} Gal, R.R., de Carvalho, R.R., Brunner,  
	R., et al. 2000 AJ, 120, 540  
 \bibitem[Gioia \& Luppino (1994)]{GL94} Gioia I.M. \& Luppino G.A. 1994,  
	AJS, 94, 5838   
 \bibitem[Kepner et al. 1999]{Kep} Kepner, J., Fan, X., Bahcall, N., et al.
	1999 ApJ, 517, 78
 \bibitem[Kim et al. 2000]{Kim}Kim, R., Strauss, M., Bahcall, N., et al. 
	2000, in {\it Clustering at High Redshift}, ASP Conference Series, 
	Vol. 200, p.422 
	Edited by A. Mazure, O. Le Fevre, and V. Le Brun. 
 \bibitem[Lumsen et al. 1992]{Lum92} Lumsden, S.L., Nichol, R.C., Collins,  
	C.A., et al. 1992, MNRAS, 258, 1  
 \bibitem[Olsen et al. 1999]{Olsen99} Olsen, L.F., Scodeggio, M., da  
	Costa, L., et al. 1999, A\&A, 345, 6810  
 \bibitem[Postman et al. 1986]{Post86} Postman, M., Geller, M.J., Huchra,  
	J.P. 1986, AJ, 91, 1267  
 \bibitem[Postman et al. 1996]{Post96} Postman, M., Lubin, L.M., Gunn,  
	J.E., et al. 1996, AJ, 111, 615  
 \bibitem[Puddu et al. 2000]{memsait} Puddu, E., Andreon, S., Longo, G.,  
        et al. 2000, MmSAI, vol. 71, n.4, in press  
 \bibitem[Schectman 1985]{Schec85} Schectman, S.A. 1985, AJS, 57, 77  
 \bibitem[Schneider, Gunn \& Hoessel (1983)]{SGH83} Schneider, D.P., Gunn,  
	J.E., Hoessel, J. 1983, ApJ, 264, 337  
 \bibitem[Stanford et al. 1998]{Stan98} Stanford, S.A., Eisenhardt, P.R.   
	\& Dickinson, M. 1998, 1998, ApJ, 492, 461  
 \bibitem[Strazzullo 2001]{thver} Strazzullo, V., Master Thesis, in preparation  
 \bibitem[Struble \& Rood 1999]{Strub99} Struble, M.F. \& Rood, H.J. 1999,  
	ApJS, 125, 35  
 \bibitem[Sutherland 1988]{Suth88} Sutherland, W. 1988, MNRAS, 234, 159  
 \bibitem[Thuan \& Gunn 1976]{TG76} Thuan, T.X. \& Gunn, J.E. 1976,  
	PASP, 88, 543  
 \bibitem[Visvanathan \& Sandage 1977]{VS77} Visvanathan, N. \& Sandage, A. 1977,  
	ApJ, 216, 214  
 \bibitem[Wade, Hoessel \& Elias 1979]{Wade} Wade, R.A., Hoessel, J.G., Elias, J.H.  
	et al. 1979, PASP, 91, 35  
 \bibitem[Weir, Djorgovski \& Fayyad 1995]{Weir95a} Weir, N., Djorgovski,  
	S.G., Fayyad, U.M. 1995, AJ, 110, 1  
 \bibitem[Weir et al. (1995)]{Weir95b} Weir, N., Fayyad, U.M., Djorgovski,  
	S.G., Roden, J. 1995, PASP, 107, 1243  
 \bibitem[Zwicky et al. 1961-68]{Zwi} Zwicky, F., Herzog, E., Wild, P., et  
	al. 1961-68, {\it Catalogue of Galaxies \& Clusters of Galaxies}.  
  
\end{thebibliography}
\end{document}